\documentclass{article}              % DO NOT DELETE THIS LINE
\usepackage{authblk}
\usepackage{siunitx}
\usepackage{graphicx}
\begin{document}                  % DO NOT DELETE THIS LINE

\title{Evaluation of modified uniformly redundant arrays as structured illuminations for ptychography}

\author[a]{Daniel J. Ching}
\author[a]{Selin Aslan}
\author[b]{Viktor Nikitin} 
\author[a]{Michael J. Wojcik}
\author[a,c]{Do\u ga G\" ursoy}

\affil[a]{X-ray Science Division, Argonne National Laboratory, 9700 Cass Avenue, Lemont, Illinois 60439, {USA}}
\affil[b]{MAX IV Laboratory, Fotongatan 2, 225 92 Lund, {Sweden}}
\affil[c]{Department of Electrical and Computer Engineering, Northwestern University, 2145 Sheridan Road, Evanston, Illinois 60208, {USA}} 

% \keyword{ptychography}
% \keyword{computational imaging}
% \keyword{structured illumination}
% \keyword{modified uniformly redundant array}

\maketitle                        % DO NOT DELETE THIS LINE

% \begin{synopsis}
% Modified uniformly redundant array provide superior convergence and spatial frequency recovery when used as the illumination pattern for ptychography when compared with conventional defocused Fresnel zone plates.
% \end{synopsis}

\begin{abstract}
Previous studies have shown that the frequency content of an illumination affects the convergence rate and reconstruction quality of ptychographic reconstructions.
In this numerical study, we demonstrate that structuring a large illumination as a modified uniformly redundant array (MURA) can yield higher resolution and faster convergence for ptychography by improving the signal-to-noise ratio of high spatial frequencies in the far-field diffraction pattern.
\end{abstract}

\section{Introduction}
% This paragraph answers the question "Why now?"
Thanks to the ongoing development of coherent light sources at short wavelengths (\SI{10}{\pico\meter}) and immense computational infrastructure, ptychography~\cite{Pfeiffer2018} has become a desirable lensless imaging technique that can provide diffraction-limited spatial resolution with no lens-imposed limitations.
Particularly for hard x-ray microscopy, lensless imaging means avoiding the low flux characteristics of optics that are not efficient refractors at high photon energies.
Lensless imaging also avoids other disadvantages of x-ray optics such as their limitation on the depth of field or their high cost.

% This answers the question "Why do other people use structured illumination"
Previous studies have shown that the structure of the illumination wavefront influences both the resolution of the reconstructed image and the convergence rate of iterative ptychography solvers \cite{Guizar-Sicairos2012, Maiden2013, Li2016, Odstrcil2019}.
The mechanism by which the resolution is affected is a decrease in the dynamic range of the far-field diffraction pattern.
Since the photon counts at the detector are described as a Poisson process, areas of the detector which record lower counts (high frequencies areas) are much more affected by noise.
Thus, illumination fields that reduce the dynamic range of the far-field diffraction pattern (i.e. help distribute photon counts to high frequency areas of the detector) will improve the signal to noise ratios for high frequencies.

For ptychography, the main way to add structure to the illumination seems to be either defocusing a zone plate or creating a random illumination with a pinhole array \cite{Maiden2011, Maiden2013}, a diffuser \cite{Stockmar2013}, or a special zone plate \cite{Marchesini2016, Morrison2018}.
However, other imaging modalities using structured illumination have used other intentional patterns such as parallel slits at different angles \cite{Gustafsson2000, Chang2009}, uniformly redundant arrays \cite{Fenimore1978, Olmos:92}, and maximum length sequences \cite{Asif2015, Boominathan2016}.

Recently, a manufacturing strategy was proposed~\cite{Marchesini:19} for manufacturing zone plates which are capable of providing arbitrary illumination structure in the focal plane.
So in this paper, we provide numerical analysis of structured illuminations from the class of sequences known as modified uniformly redundant arrays (MURAs) and quantitatively compare their ptychographic performance with the currently utilized illuminations for ptychography.

\section{Structured Illuminations}

\begin{figure}
\centering
\includegraphics[width=\textwidth]{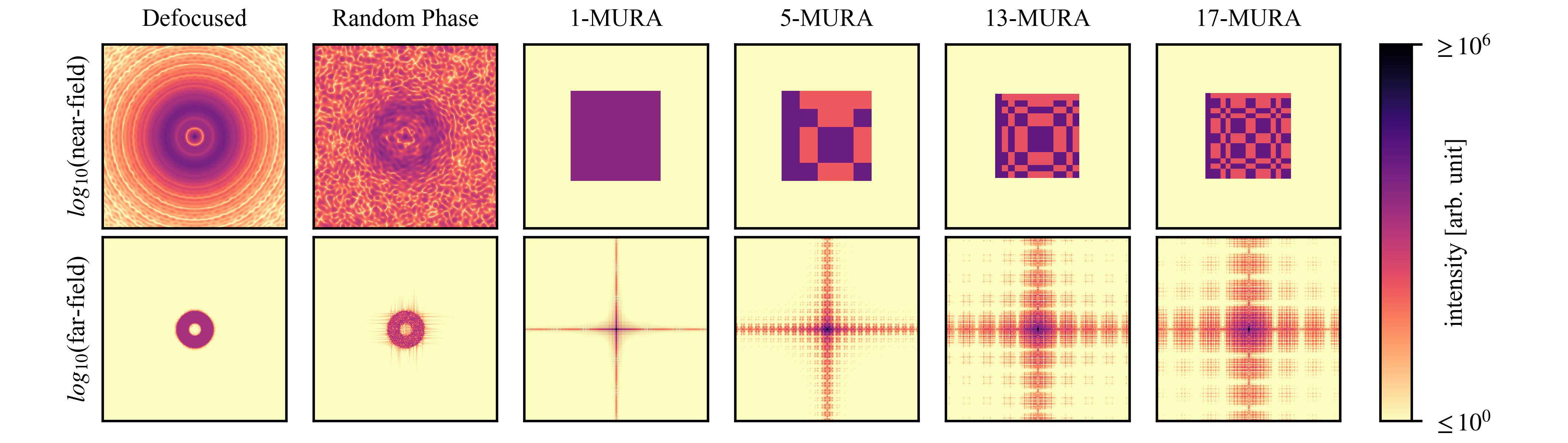}
\caption{A comparison of near-field and far-field intensities (phase not shown) for some illuminations.}
\label{fig:fft_compare}
\end{figure}

Our study compares a defocused zone plate illumination, a randomized-phase defocused zone plate illumination, and many modified uniformly redundant array (MURA) illuminations of different resolutions.
These illuminations, some of whose near and far-field intensities are shown in Fig.~\ref{fig:fft_compare}, have an approximate \SI{1200}{\nano\meter} width (the widths of the MURAs vary slightly so that the MURA width is a multiple of the reconstruction grid pixel size), but do not have the same footprint, so their intensities are normalized to have the same total photon flux.
The pixel size for the near-field illustrations and reconstruction grid is \SI{10}{\nano\meter}.

The zone plate illuminations are based on a simulation of the Fresnel zone plate used at the Velociprobe~\cite{Deng2019} at the Advanced Photon Source at the U.S. Department of Energy's Argonne National Laboratory (Lemont, IL, USA).
The defocused illumination represents the current method for increasing the size of the illumination spot and adding ``structure" to the illumination.
This method is convenient because it only requires moving the object of interest beyond the focal plane. 
The random-phase illumination is a more complex method of adding structure to the illumination, and may be accomplished by adding a diffuser in the beam path or by using a specially designed zone plate whose ring are randomly offset to provide a random phase modulation.
The diffuser could be a plate with random holes in it~\cite{Maiden2011, Maiden2013}, or it could be a material that adds a spatially-random phase shift to the wavefront~\cite{Stockmar2013}.
We simulated a randomized phase for the defocused beam by randomizing the phase of the wavefront at the focus.
This random-phase illumination represents the current best method for reducing the dynamic range of the far-field diffraction pattern without using a beam stop.

The rest of the illuminations are modified uniformly redundant arrays (MURAs) \cite{Gottesman1989}.
MURAs are 1D binary sequences whose lengths $L$ are, by definition, prime numbers that can be written in the form of $L = 4n + 1$, where $n$ is a positive integer.
The $i$ th bit in the sequence is $1$ if $i$ is a quadratic residue modulo $L$; otherwise it is $-1$.
Two-dimensional MURAs are generated by array-multiplying one-dimensional MURAs.
Our simulated MURA illuminations are 0\% or 90\% absorbing when the MURA is -1 or 1 respectively.
Each MURA illumination is labeled n-MURA, where n is the number of elements in the 1D MURA used to generate it.
The 1-MURA is merely a square illumination with no internal structure.
Some MURA illuminations are marked with an asterisk (*) indicating that they have features smaller than \SI{30}{\nano\meter} in the focal plane.

Specially structured illuminations such as the MURA or others could be created using multiple methods.
First, they could be created using specially manufactured zone plates such as those proposed by \cite{Marchesini:19} which are designed to have a specific pattern at the focal plane.
Second, they could be created by placing a spatially coded aperture in the beam path before the specimen.
Metal-assisted chemical etching of Si has produced submicron lithographically generated structures with aspect ratios of several hundreds \cite{Li2017, Chang2014} suitable for the spatially coded aperture dimensional requirements.
Another option is to manufacture the coded aperture by deep reactive ion etching of Si followed by electroplating Au into the resulting mold \cite{Hollowell2019} which would have been lithographically patterned.

\section{Effects of structured illumination on far-field illumination}

\begin{figure}[]
\centering
\includegraphics{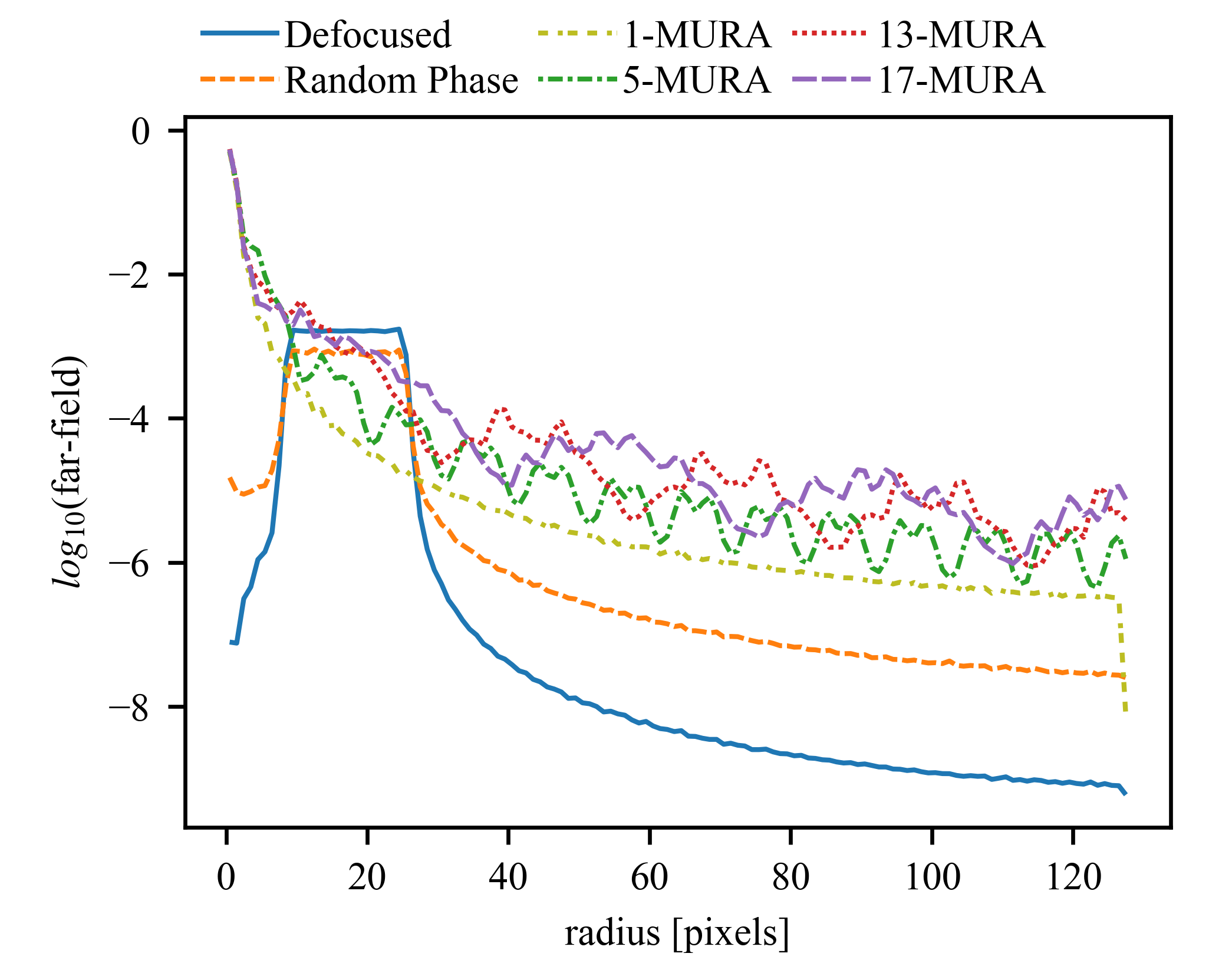}
\caption{The average far-field intensity as a function of radius for some illuminations.}
\label{fig:fft_range}
\end{figure}

The ability of structured illuminations to reduce the dynamic range of the far-field diffraction pattern and improve its signal-to-noise characteristics is what makes them useful.
Fig.~\ref{fig:fft_range} shows the average far-field intensity as a function of distance from the center (radius) of the far-field diffraction patterns for each illumination.
As shown in previous studies~\cite{Maiden2013} and in Fig.~\ref{fig:fft_range}, randomizing the phase of a conventional illumination does decrease the dynamic range of the far-field diffraction pattern.
In our simulation, the random-phase zone plates the intensity spans approximately four orders of magnitude compared to the defocused and MURA illuminations which span approximately 6 order of magnitude.

However, the MURA illuminations have a higher on average far-field intensity than the zone plate illuminations, and increasing the resolution of the MURA only increases the average far-field intensity.
This is important because our illuminations are all flux-normalized.
This means that for a given noise level and illumination size, high-frequency information about the object (which is located at the edges of the detector) is more likely to be above the noise floor for the MURA illuminations than it is for zone plate illuminations.
In other words, the total flux on the sample must increase to take advantage of the lower dynamic range of the random-phase illumination. 

\section{Effects of structured illumination on reconstruction quality}

\begin{figure}
\centering
\includegraphics{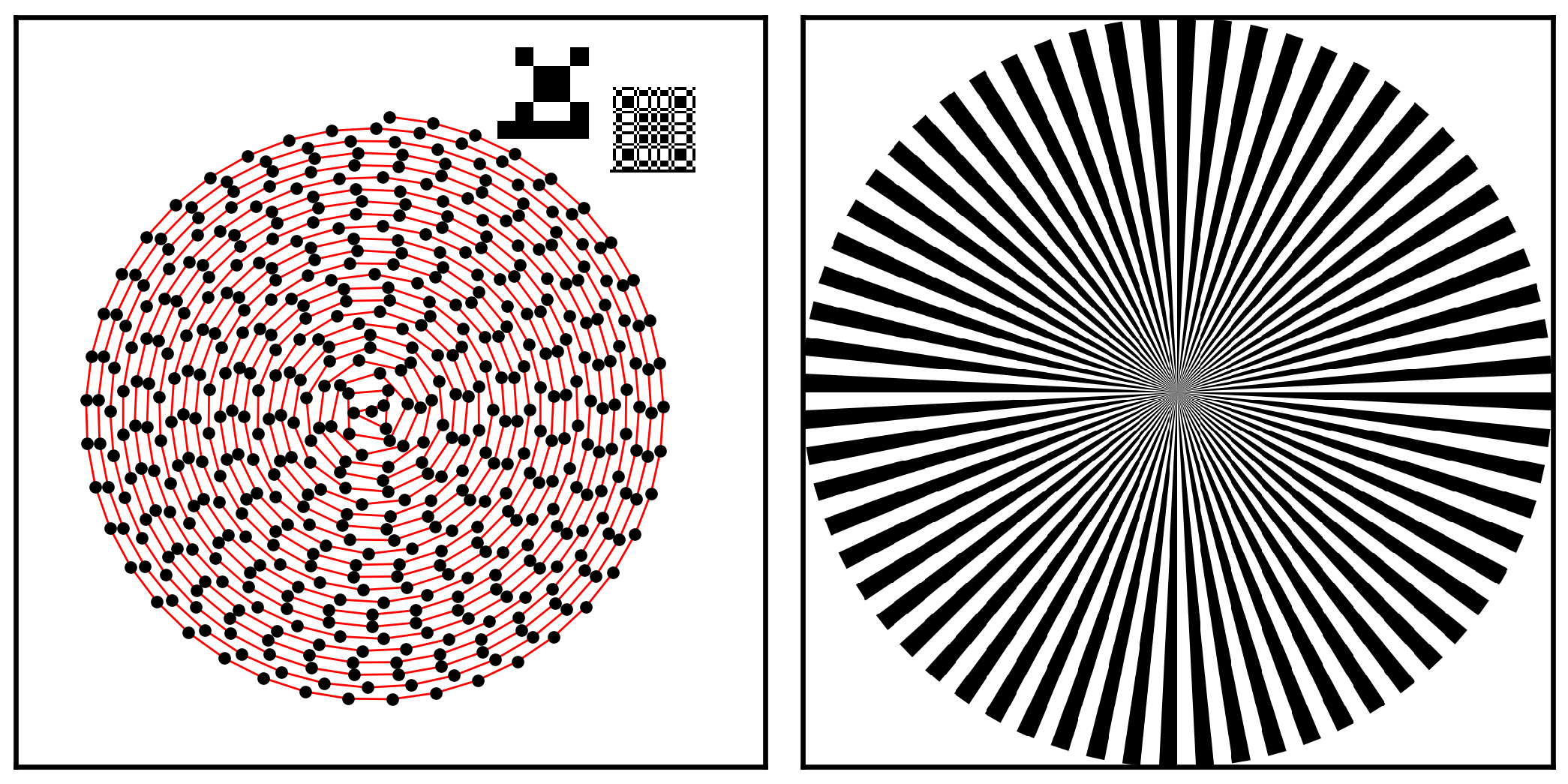}
\caption{(left) Scanning positions and MURA spot size. (right) The phase component of the simulated phase-only phantom.}
\label{fig:trajectory}
\end{figure}

We simulated the data acquisition using each of the illuminations for 20 trials with Poisson noise.
Our simulated phase-only phantom was a 1024 by 1024 pixels with a phase that was a Siemens Star with 64 spokes.
The star is a binary phase object whose spokes cause a phase offset of one.
The step-scanning trajectory was 512 frames collected in a spiral pattern of constant linear velocity.
These are shown in Fig.~\ref{fig:trajectory}.
The probe and detector grids were 256 by 256 pixels; this meets the criterion that the detector should be at least twice the size of the illumination diameter in pixels since none of our illuminations are wider than 128 pixels at full width half maximum \cite{Jacobsen2017}.

To reconstruct the amplitude and the phase of the phantom from the far-field measurements, we solved the following least-square minimization problem,
$$
\||\mathcal{F}\{P\cdot O\}|-\sqrt{D}\|_2\to \text{min},
$$
with the nonlinear conjugate-gradient method~\cite{DaiYuan:99}.

\begin{figure}
\centering
\includegraphics{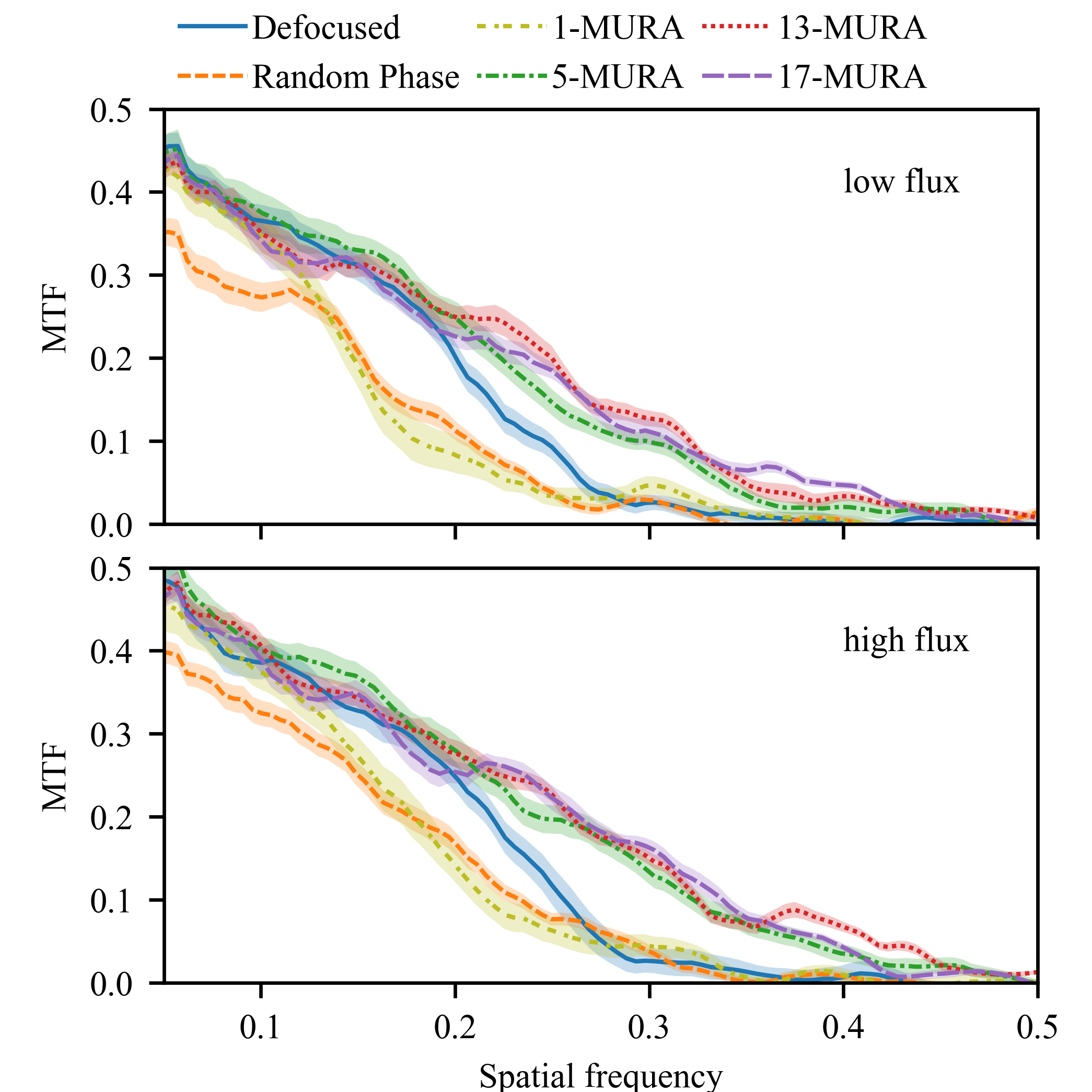}
\caption{Comparison of the modulation transfer function (MTF) of reconstructed Siemens stars using conventional zone-plate and MURA illuminations. Line shadows show one standard deviation from 20 noisy trials.}
\label{fig:mtf}
\end{figure}

To quantitatively compare the ptychography reconstructions, we computed the modulation transfer function (MTF) from the Siemens star by fitting a periodic function to the spokes of the star to measure the amplitude of the reconstructed at frequencies above the Nyquist limit \cite{loebich2007}.
This method is implemented in XDesign~\cite{Ching:17}.
The MTF for the converged reconstructions of the zone plate illuminations and MURAs with resolutions 1...17 is shown in show in Fig.~\ref{fig:mtf} for a high and low noise level.
These MTF plots and all the others have been smoothed with a second order Savitzky-Golay filter to make them more decipherable.
% As discussed in previous studies~\cite{Guizar-Sicairos2012, Maiden2013, Li2016, Odstrcil2019}, structured illuminations with a flatter frequency response have faster convergence rates and better reconstruction quality at high noise levels.
We can see that the MURA illuminations with resolution greater than one (the illuminations with actual structure) are better than the zone plate based illuminations.
The MURA illumination MTFs have higher values at the short spatial frequencies.
This means the MURAs are capable of preserving the short spatial frequency information and will produce better reconstructions.

In our flux-normalized case, the random-phase illumination, performs worse than the constant-phase defocused illumination.
Referring back to the photon distribution on the detector (Fig.~\ref{fig:fft_range}), the random-phase illumination has a smaller dynamic range, but it also has a lower peak intensity than the defocused illumination.
This means that for any noise level where the high frequency information of the random-phase illumination is below the noise floor, the defocused illumination actually provides better signal to noise ratios and provides information about the same frequencies.
Regardless, the MURA illuminations always have more frequencies above the noise floor.

\begin{figure}
\centering
\includegraphics{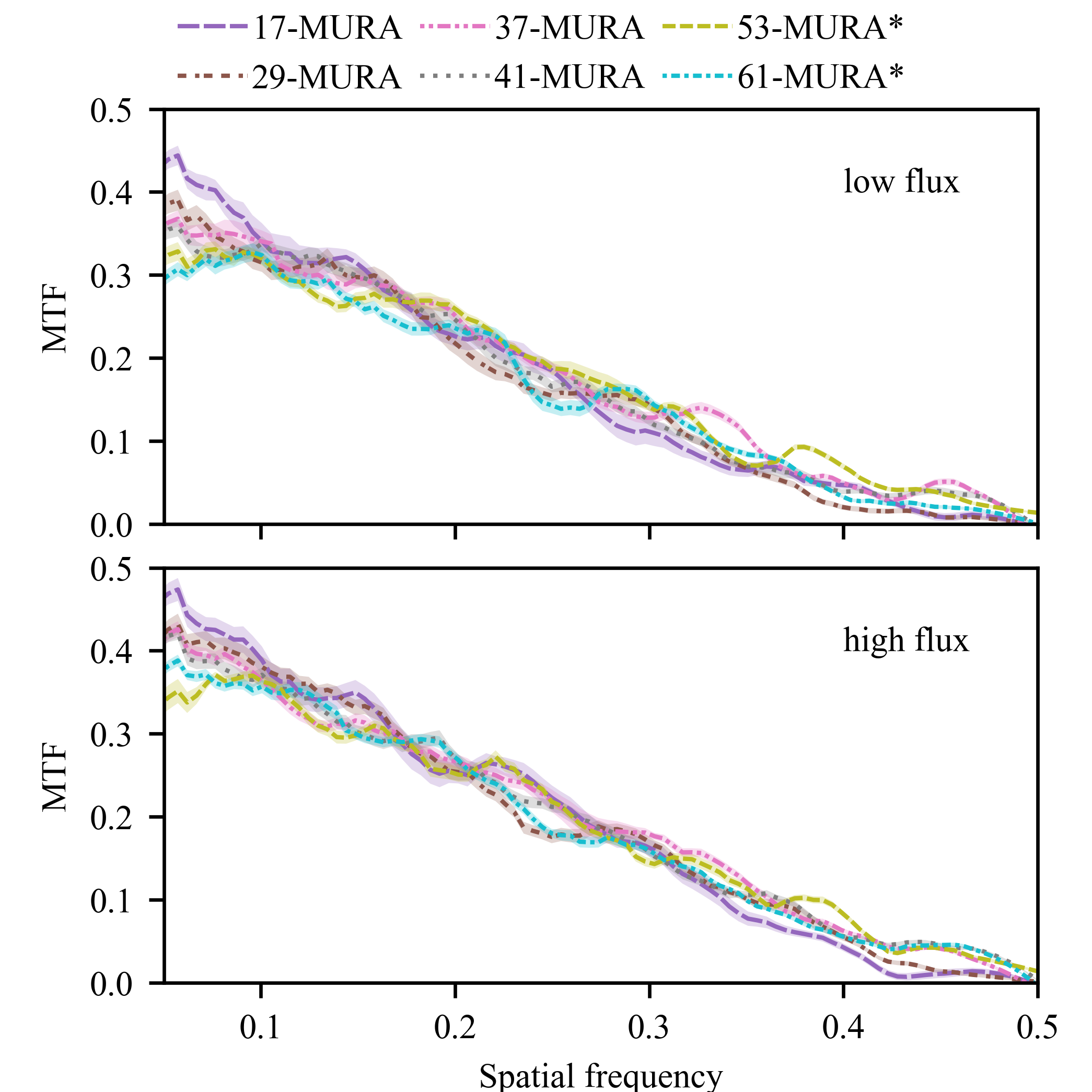}
\caption{Comparison of the modulation transfer function of reconstructed Siemens stars for MURAs of many resolutions. Line shadows show one standard deviation from 20 noisy trials.}
\label{fig:mtf-sizes}
\end{figure}

We included additional MURA resolutions in our study because we wanted to see the effect that increasing the MURA resolution had on the MTF of the Siemens stars.
These MTFs for high and low noise levels are shown in Fig.~\ref{fig:mtf-sizes}.
In this figure, MURAs marked with an asterisk (*) are have features smaller the \SI{30}{\nano\meter} (3 pixels) which is smaller than what we consider manufacturable.
From this data, we see that using higher resolution MURAs may slightly improve the recovery of very short spatial frequencies.
This agrees with the single order of magnitude frequency response increases that we observe in Fig.~\ref{fig:fft_range}.

\begin{figure}
\centering
\includegraphics{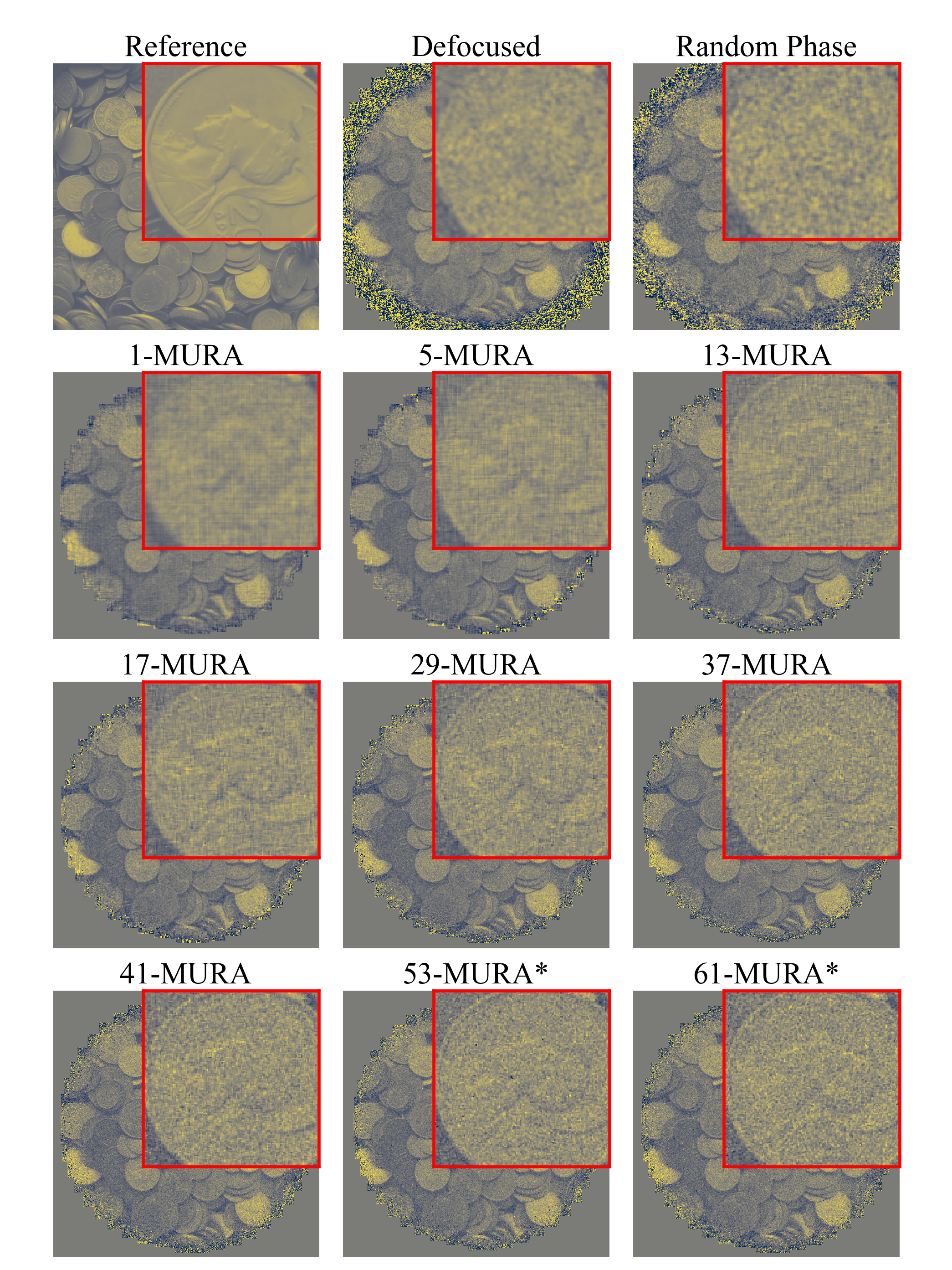}
\caption{Reconstructions at the low noise level using a image of coins.}
\label{fig:coins}
\end{figure}

To show what these modest increases in fidelity may look like we provide example reconstructions using the coins image~\cite{wiki:coins} shown in Fig.~\ref{fig:coins} for comparing nine different MURA resolutions.
Here we can see that the higher resolution MURAs reconstructions appear slightly sharper features at the cost of increased noise.

% \begin{figure}
% \centering
% \includegraphics{mtf-mura-attenuation}
% \caption{The modulation transfer function of reconstructed Siemens stars with low noise (top) and high noise (bottom). Line shadows show one standard deviation from 20 noisy trials.}
% % \label{fig:mtf-attenuation}
% \end{figure}

The reader may also be curious about the effect of decreasing the attenuation of the dark patches of the MURA illumination. The MURAs in this study are 90\% absorbing in the ``dark" pixels of the MURA pattern.
Decreasing the absorption of these dark patches would smoothly transition a given MURA to have the properties of the 1-MURA of the same width.

\begin{figure}
\centering
\includegraphics{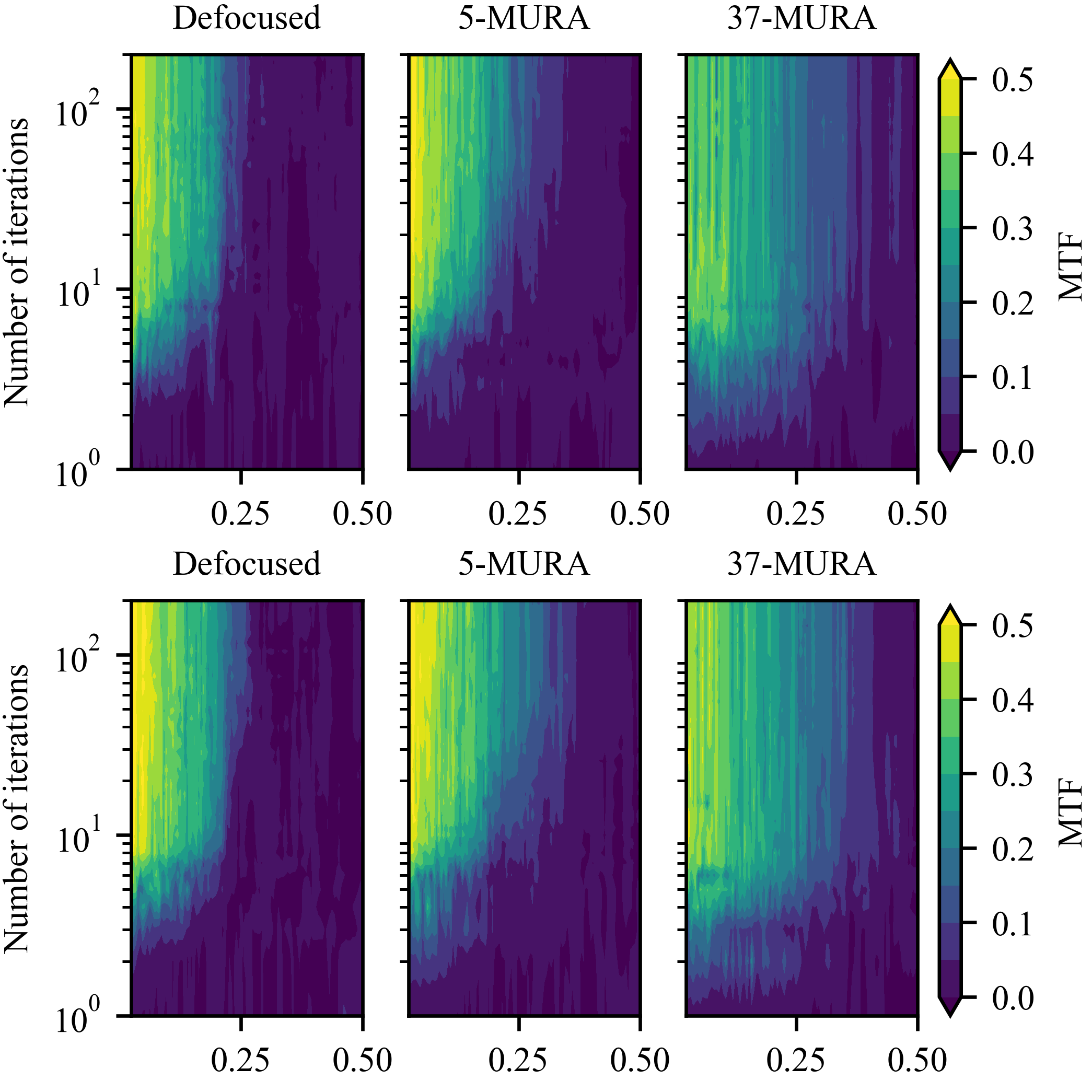}
\caption{Comparison of the modulation transfer function convergence rates of the different illuminations for high and low noise.}
\label{fig:mtf-convergence}
\end{figure}

We also compared the convergence rates of our reconstruction algorithm using different illuminations.
The MTFs for three of the illuminations as a function of the number of completed conjugate gradient iterations are shown in Fig.~\ref{fig:mtf-convergence}. Here we see that the MURA illuminations of high resolution not only converge to better reconstructed images, but they also do it in less iterations.

\section{Conclusions}

In this letter, we have shown that we can theoretically do better than randomized-phase zone plates for structured illumination in ptychography by using intentionally designed illuminations such as the MURA.
The reasons for this is are that for a given amount of flux, the MURA illumination puts more photons into the short spatial frequency components of the far-field diffraction pattern.
This leads to much better noise robustness at high spatial frequencies than for a randomized-phase illumination.
Also, convergence rates for MURA structured probes are faster than for random-phase illuminations when the resolution of the MURA is sufficiently high.

\section*{Acknowledgements}
This material is based upon work supported by the U.S. Department of Energy under Contract No. DE-AC02-06CH11357.
% U.S. Department of Energy (DE-AC02-06CH11357);
The research is also supported by the Swedish Research Council (2017-00583).
% Swedish Research Council (2017-00583).

\bibliography{main}

\end{document}